\documentclass[10pt,twocolumn,brazil,english,amsmath,amssymb, aps]{revtex4-2}
\pdfoutput=1
\usepackage[T1]{fontenc}
\usepackage[latin9]{inputenc}
\usepackage{babel}
\usepackage{float}
\usepackage{textcomp}
\usepackage{url}
\usepackage{graphicx}
\usepackage{hyperref}
\usepackage{soul}

\makeatletter

\providecommand{\tabularnewline}{\\}

\makeatother

\begin{document}
\title{Can dark energy explain a high growth index?}
\author{Ícaro B. S. Cortês}
\email{icaro.cortes.710@ufrn.edu.br}
\affiliation{Departamento de Engenharia de Computação e Automação, Universidade
Federal do Rio Grande do Norte, Caixa Postal 1524, CEP 59078-970, Natal, Rio
Grande do Norte, Brazil.}

\author{Ronaldo C. Batista}
\affiliation{Escola de Ciências e Tecnologia, Universidade Federal do Rio Grande
do Norte, Caixa Postal 1524, CEP 59078-970, Natal, Rio Grande do Norte, Brazil.}

\date{\today}

\begin{abstract}
A promising way to test the physics of the accelerated expansion of the
Universe is by studying the growth rate of matter fluctuations, which can be
parametrized  by the matter energy density parameter to the power $\gamma$, the
so-called growth index. It is well-known that the $\Lambda$CDM cosmology
predicts $\gamma=0.55$. However, using observational data, Ref.
\citep{Nguyen:2023fip} measured a much higher
$\gamma=0.633^{+0.025}_{-0.024}$, excluding the $\Lambda$CDM
value within $3.7\sigma$. In this work, we analyze whether Dark Energy (DE) with
the Equation of State (EoS) parameter described by the CPL parametrization can
significantly modify $\gamma$ with respect to the predicted $\Lambda$CDM one. Besides the usual Smooth DE (SDE) scenario, where DE perturbations are neglected on small
scales, we also consider the case of Clustering Dark Energy (CDE), which has
more potential to impact the growth of matter perturbations. In order to
minimally constrain the background evolution and assess the largest meaningful
$\gamma$ distribution, we use data from $32$ Cosmic Chronometers, $H(z$), data
points. In this context, we found that both SDE and CDE models described by the
CPL parametrization have almost negligible probability of providing $\gamma>0.6$.
However, given that the measured $\gamma$ value assumes the $\Lambda$CDM background, a direct statistical measure of the incompatibility between theory and the measured value can not be done for other backgrounds. Thus, we devise a method in order to make a quick estimation of the $\gamma$ constraints for CPL background. This method indicates that, when using DESI DR2 BAO data to constrain background parameters, no significant changes in the $\gamma$ central value and uncertainty is observed. Consequently, both SDE and CDE described by CPL parametrization shall have a comparable level of tension with the expected $\gamma$ measurements. Moreover, we present new fitting functions for $\gamma$, which are more accurate and general than the one proposed in Ref. \citep{Linder:2005in} for SDE, and, for the first time, fitting functions for CDE models.
\end{abstract}

\maketitle
\section{Introduction}

The accelerated expansion of the Universe is still a big question
in Cosmology. It can be explained either by a modified theory of gravity,
or a unknown component of the universe, labeled as Dark Energy (DE)
\citep{Abdalla:2022yfr}. In the pursuit of more accurate data to answer
this question, some tensions arose. The most significant one is known
as the Hubble tension, which reflects the difference between measurements
of the present expansion rate, $H_{0}$, obtained locally by using
the distance ladder methods \citep{Riess2021} and globally, assuming
the $\Lambda$CDM model, from Cosmic Microwave Background (CMB) data
\citep{Aghanim2020}. There is also the $S_{8}$ tension (where
$S_{8}\equiv\sigma_{8}\sqrt{\Omega_{m0}/0.3}$,
$\sigma_{8}$ is the amplitude of matter fluctuations at $8h^{-1}\text{Mpc}$
and $\Omega_{m0}$ is the matter density parameter now), which is
related to the difference between the values of these parameters inferred
from CMB \citep{Aghanim2020} and measurements of galaxy clustering
and weak gravitational lensing, as discussed in \citep{DiValentino2021}. 

The $S_{8}$ tension is directly related to the growth of cosmic perturbations and has been analysed frequently in the literature, e.g., \cite{Nunes:2021ipq,Adil:2023jtu,ACT:2023kun}. A particularly simple and promising way to test how the physics of the accelerated expansion impacts the evolution of matter perturbations
is by analyzing the growth rate of matter perturbations \citep{Amendola2013}
\begin{equation}
f=\frac{d\ln\delta_{m}}{d\ln a}\,,\label{eq:growth-rate}
\end{equation}
which can be parametrized as \citep{Linder:2005in}
\begin{equation}
f\simeq\Omega_{m}^{\gamma}\,,
\end{equation}
where $\gamma$ is a constant that depends weakly on the Equation
of State (EoS) parameter of Smooth DE (SDE) models,
$w\left(t\right)=p_{de}\left(t\right)/\rho_{de}\left(t\right)$,
and $\Omega_{m}=\Omega_{m}\left(a\right)$ is the matter energy density
parameter. For the $\Lambda$CDM cosmology, Refs.
\citep{Wang:1998gt,Linder:2005in}
found $\gamma=0.55$. However, using observational data, Ref.
\citep{Nguyen:2023fip}
found a much higher $\gamma=0.633$ (implying a suppression of growth),
excluding the $\Lambda$CDM value by $3.7\sigma$, which can be referred as `$\gamma$ tension'. In the same work,
it was also shown that a higher $\gamma$ value reduces the $S_{8}$
tension from $3.2\sigma$ to $0.9\sigma$.

This high value of $\gamma$ naturally raises the question of what
theoretical mechanisms are capable of producing it. In the case of
SDE, it was shown that $\gamma$ can be accurately described by the
fitting function \citep{Linder:2005in},
\begin{equation}
\ensuremath{\gamma(w_{1})=\begin{cases}
0.55+0.02(1+w_{1}), & w_{1}\le-1\\
0.55+0.05(1+w_{1}), & w_{1}>-1
\end{cases}},\label{eq:gamma-linder}
\end{equation}
where $w_{1}=w\left(z=1\right)$. This indicates that, for $w_{1}\le-1$,
we would need $w_{1}=3.15$ to get $\gamma=0.633$, in contradiction
with the parametrization condition. For $w_{1}>-1$, we would need
$w_{1}=0.66$, which would generate a large DE density around $z=1$.
This simple extrapolation exercise indicates that it should be very
difficult for SDE models to produce a high growth index. As we will
show, with very loose constraints on the background evolution based
on Cosmic Chronometers (CC) data, SDE with CPL parametrization has
an almost negligible probability of producing $\gamma>0.6$.

The usual assumption of SDE, in which one usually neglects DE perturbations
on small scales, is based on Quintessence models
\citep{Peebles:1987ek,Ratra:1987rm,Wetterich:1987fm}.
In this case, DE is described by a minimally coupled canonical scalar
field. The linear perturbations of this field propagate with sound
speed $c_{s}=1$, not allowing its perturbations to grow significantly
on small scales. This approximation is well justified even in the nonlinear
regime \cite{Batista:2022ixz}. The simplest generalization of this scenario can
be done by describing DE as a non-canonical minimally coupled scalar
field, called k-essence \citep{ArmendarizPicon1999,ArmendarizPicon:2000ah}.
In this case, $c_{s}$ can be chosen and DE perturbations can grow
at small scales, see \citep{Batista:2021uhb} for a more detailed
discussion. In this work, we will consider the limiting case of
$c_{s}\rightarrow0$,
in which DE perturbations have the maximal potential to grow and impact
the evolution of matter perturbations. We refer to this scenario as Clustering
DE (CDE), which growth index has already been studied in
\citep{Batista2014a,Mehrabi2015,Mehrabi2015a,Mehrabi2015b}.

In this paper, we explore how SDE and CDE can change the growth index. We
confirm the expectation based on the fitting formula (\ref{eq:gamma-linder})
that SDE model can not provide $\gamma > 0.6$. We also find
that CDE models can not raise the values of $\gamma$ significantly.

As we will show, for the case of CDE, this result has a simple explanation.
When $w\left(t\right)>-1$, DE perturbations have the tendency
of being correlated with matter perturbations ($\delta_{de}\propto\delta_{m}$)
and anti-correlated for $w\left(t\right)<-1$ ($\delta_{de}\propto-\delta_{m}$).
In order to raise $\gamma$, DE perturbations must be anti-correlated
with $\delta_m$, causing a decrease in the gravitational potential and
consequently slowing down the growth rate (higher $\gamma$). However, in the
case $w\left(t\right)<-1$, the DE energy density decreases rapidly at
high-$z$, thus the overall impact is very limited. On the other hand,
the case $w\left(t\right)>-1$ can easily enhance the matter growth,
thus providing significantly lower $\gamma$ values, as already
shown in \citep{Batista2014a}.

Based on the samples of $\gamma$ that we have computed using CC data, we were
able to construct a new parametrization $\gamma=\gamma\left(w_{1}\right)$,
which is more accurate than (\ref{eq:gamma-linder}) and valid for
larger parameter space. Moreover, for the first time, we present a $\gamma$
parametrization for the case of CDE.

We also perform a more specific analysis using DESI DR2 Baryon Acoustic Oscillations (BAO) data, Ref. \cite{DESI:2025zgx}, plus CMB based priors to constrain background parameters. In this context, we make use of synthetic growth rate function to make a fast estimation of $\gamma$ constraints beyond $\Lambda$CDM. This study suggests that the central values of $\gamma$ and its uncertainty for $w$CDM and CPL models should be similar to the ones measured for $\Lambda$CDM.

The plan of this paper is as follows. In Sect. \ref{sec:Background-cosmology},
we define the background cosmology, the data, the parameters priors
and statistics used to constrain the background evolution for the general analysis. In Sect.
\ref{sec:Growth-of-Structure},
we present the equations for the evolution of matter and DE perturbations and discuss the accuracy of the constant-$\gamma$ parametrization in describing $f(z)$.
The resulting distributions for background parameters and $\gamma$ and fitting functions are shown in Sect. \ref{sec:Results}. In Sect. \ref{estimation}, we further constrain the background parameter using DESI BAO DR2 data and present a simple procedure to make a fast estimation of the $\gamma$ constraints for models beyond $\Lambda$CDM and analyse the statistical compatibility between the theoretical values and the simulated measurements.
We conclude in Sect. \ref{sec:Conclusion}.

\section{\label{sec:Background-cosmology}Background cosmology}

In this work, we assume General Relativity and a flat universe described by the
Friedmann-Lemaître-Robertson-Walker metric, in which the line element
is represented by

\begin{equation}
ds^{2}=-c^{2}dt^{2}+a(t)^{2}\left[dr^{2}+r{}^{2}d\Omega^{2}\right],
\end{equation}
where $a(t)$ is the scale factor. As such, the square of the Hubble
function can be written as
\begin{equation}
H^{2}=\left(\frac{\dot{a}}{a}\right)^{2}
=H_{0}^{2}\left(\Omega_{m0}a^{-3}+
\Omega_{de}\left(a\right)\right)\,,
\label{eq:hubble-function}
\end{equation}
where $\Omega_{m0}$ is the matter (dark matter plus baryons) density
parameter now and  $\Omega_{de}\left(a\right)$ is the DE energy parameter, which
depends on the EoS assumed.

We consider that DE EoS is given by Chevallier-Polarski-Linder (CPL)
\citep{Chevallier:2000qy,Linder:2002et} parametrization:
\begin{equation}
w(a)=w_{0}+w_{a}(1-a).
\end{equation}
Thus, the DE density parameter is given by
\begin{equation}
\Omega_{de}(a)=(1-\Omega_{m0})a^{-3(1+w_{0}+w_{a})}\exp\left(-3w_{a}(1-a)\right)
.
\end{equation}

Besides the general case of free $w_{0}$ and $w_{a}$, we will also
analyze the $\Lambda$ limit ($w_{0}=-1$ and $w_{a}=0$) and constant
EoS case (free $w\equiv w_{0}$ and $w_{a}=0$). We refer to these
models as CPL, $\Lambda$CDM and $w$CDM, respectively. In the general
case, we have four free parameters: $h=H_{0}/\left(100\text{km
s}^{-1}\text{Mpc}^{-1}\right)$,
$\Omega_{m0}$, $w_{0}$ and $w_{a}$. Next, we discuss how to minimally
constrain these parameters using $H\left(z\right)$ data.

\subsection*{Cosmic Chronometers Data}

In order to minimally constrain the background evolution and the
model parameters, we make use 32 of the most recent CC data available,
compiled by \citep{Favale2023}. We use the Python library \texttt{Emcee}
\citep{ForemanMackey2012} as a Monte Carlo Markov Chain (MCMC) sampler
and Python library \texttt{GetDist} \citep{Lewis2019} to analyse
and plot the posteriors distributions.

There is a well-known discussion about the systematic errors in CC
data, \citep{Moresco2012,Moresco2015,Moresco2016}. In these works,
systematics of 15 of the 32 data points have been analyzed. Here,
we split the dataset in two groups (with and without systematics), and
$\chi^{2}$ given by
\begin{equation}
\chi^{2}=\chi_{{\rm nosys}}^{2}+\chi_{{\rm sys}}^{2}\,,\label{eq:chi^2}
\end{equation}
where $\chi_{{\rm nosys}}^{2}$ is associated with the 17 data points
without systematics, provided by
\citep{Zhang2014,Jimenez2003,Simon2005,Ratsimbazafy2017,Stern2010,Borghi2022},
which reads
\begin{equation}
\chi_{{\rm nosys}}^{2}=\sum_{i}\left(\frac{\Delta
H_{i}}{\sigma_{i}}\right)^{2}\,,\label{eq:chi-no-sys}
\end{equation}
where $\Delta H_{i}=H_{\text{model}}\left(z_{i}\right)-H_{{\rm
\text{data}}}\left(z_{i}\right)$
and $\sigma_{i}$ is the corresponding uncertainty of data points.
The other part of the $\chi^{2}$ associated with data points with
systematics is given by 
\begin{equation}
\chi_{{\rm
sys}}^{2}=\Delta\vec{H}^{T}\boldsymbol{Cov}^{-1}\Delta\vec{H}\,,\label{eq:chi-
sys}
\end{equation}
where $\Delta\vec{H}$ is a vector with $\Delta H_{i}$ components associated with
the 15 data points provided by Refs.
\citep{Moresco2012,Moresco2015,Moresco2016},
and $\boldsymbol{Cov}$ is the corresponding covariance matrix for the
BC3 model, available at \url{https://gitlab.com/mmoresco/CCcovariance}.

We sample the posterior distribution of the vector of parameters
$\theta=\left(h,\Omega_{m0},w_{0},w_{a}\right)$,
given by
\begin{equation}
\mathcal{L}\left(\theta|D\right)\propto
P\left(\theta\right)L\left(D|\theta\right)\,,\label{eq:posterior}
\end{equation}
where 
\begin{equation}
L\left(D|\theta\right)\propto\exp\left(-\frac{\chi^{2}}{2}\right)\label{eq:liki}
\end{equation}
is the likelihood of the data $D$ given $\theta$
and $P\left(\theta\right)$ indicates the priors assumed for the parameters,
listed in Tab. \ref{tab:Priors}. We also implement an additional
flat prior to exclude cosmologies in which
$\Omega_{de}\left(z=1000\right)>0.01$,
where $z=1/a-1$ is the redshift. This condition is similar to the
assumption $w_{0}+w_{a}<-0.1$ described by \citep{Huterer2006}.
Allowing for DE models with non-negligible energy density at high-$z$,
besides being highly disfavored by data \citep{Gomez-Valent:2021cbe},
invalidates the Einstein-de-Sitter initial conditions used to solve
the evolution of matter and DE perturbations, which will be discussed
in detail in the next section.

We will discuss the MCMC results regarding the background parameters in the section \ref{sec:Results}, together with the results for $\gamma$.

\begin{table}[h]
\caption{\label{tab:Priors}Priors for all parameters.}
\centering{}%
\begin{tabular}{cc}
Parameter & Prior\tabularnewline
\hline 
\hline 
$h$ & $\mathcal{U}[0.5,1]$\tabularnewline
$\Omega_{m0}$ & $\mathcal{U}[0.01,0.99]$\tabularnewline
\hline 
$w$ & $\mathcal{U}[-3,-1/3]$\tabularnewline
$w_{0}$ & $\mathcal{U}[-3,1]$\tabularnewline
$w_{a}$ & $\mathcal{U}[-3,3]$\tabularnewline
\hline 
\end{tabular}
\end{table}

\section{\label{sec:Growth-of-Structure}Growth of Cosmological Perturbations}

In a universe with SDE, the linear evolution of the matter density
contrast, $\delta_{m}\equiv\rho_{m}/\overline{\rho}_{m}-1$, is solely
affected by the background expansion, as described by the equation
\begin{equation}
\delta_{m}^{\prime\prime}+\left(2-\frac{1+3w\left(a\right)\Omega_{de}
\left(a
\right)}{2}\right)\frac{\delta_{m}^{\prime}}{a}-\frac{3}{2}\frac{\Omega_{
m}\left( a\right)}{a^{2}}\delta_{m}=0,
\label{eq:delta-smooth-DE}
\end{equation}
where the prime denotes a derivative with respect to the scale factor.
Starting the integration at a high redshift, $z_{i}=1000$, initial
values can be computed using the analytical solution for a matter-dominated
Einstein-de-Sitter (EdS) universe $\delta_{mi}^{\prime}=\delta_{mi}/a_{i}$,
where we assumed $\delta_{mi}>0$.

In the case of CDE, $\delta_{m}$ is also affected by DE perturbations,
$\delta_{de}$, whose effect is maximized for DE models with negligible sound
speed. In this work, we consider the extreme case $c_{s}=0$ based on the
fluid description of DE and phenomenologically allowing for an EoS which
can be non-phantom ($w(t)>-1$), phantom ($w(t)<-1$) or transit between
these regimes. In this context, matter and DE perturbations obey the
following system of equations \citep{Batista:2021uhb}:

\begin{equation}
\begin{cases}
\delta_{m}^{\prime}+\frac{q}{a^{2}} & =0\,,\\
\delta_{de}^{\prime}-3\frac{w}{a}\delta_{de}+(1+w)\frac{q}{a^{2}} & =0\,,\\
q^{\prime}-\frac{1}{2}(1+3w\Omega_{de})\frac{q}{a}+\frac{3}{2}(\Omega_{m}\delta_
{m}+\Omega_{de}\delta_{de}) & =0\,,
\end{cases}\label{eq:dm-de-system}
\end{equation}
where $q\equiv\theta/H$ and $\theta=\vec{\nabla}\cdot\vec{v}$ is
the divergence of the peculiar velocity of the DE-matter fluid. Note
that the equation for the peculiar velocity is the same for matter
and CDE because we are considering $c_{s}=0$. To compute the initial
conditions for DE perturbations, we use the solution valid for constant
$w$ in matter-dominated era
\citep{Abramo2009,Sapone2009,Creminelli2009,Batista:2013oca}

\begin{equation}
\delta_{de}=\frac{1+w}{1-3w}\delta_{m}.\label{eq:delta_de-delta_m}
\end{equation}

Although solution (\ref{eq:delta_de-delta_m}) only gives a general
qualitative behavior at low-$z$ \citep{Batista:2013oca}, it can
help us to predict the influence of CDE on the growth of $\delta_{m}$.
When $w\left(a\right)<-1$ we have the tendency $\delta_{de}\propto-\delta_{m}$,
while for $w\left(a\right)>-1$ we have $\delta_{de}\propto\delta_{m}$.
If the EoS is always phantom or non-phanton, these relations are always
valid. In the case that the EoS transits between phantom and non-phanton or
vice-versa, the actual relation between $\delta_{de}$ and $\delta_{m}$ might
take some time to achieve the expected behavior. The general trend
is clear: phantom EoS will induce negative $\delta_{de}$, lowering
the clustering power encoded in the last term of Eq. (\ref{eq:dm-de-system}),
whereas non-phantom EoS enhances growth. Therefore, the desired higher
values of $\gamma$ could be obtained in phantom CDE models.

We solve Eqs. (\ref{eq:delta-smooth-DE}) and (\ref{eq:dm-de-system}),
then determine the growth rate $f=d\ln\delta_{m}/d\ln a$ and fit
a constant $\gamma$ assuming $f=\Omega_{m}^{\gamma}(a)$ for all
the posterior distribution obtained. In this
process, it is important
to ask whether the fitted values of $\gamma$ can accurately describe
$f$. In the paper \citep{Linder:2005in}, where the fit (\ref{eq:gamma-linder})
was introduced for the SDE model described by CPL parametrization, the accuracy
of the fit was reported with respect to the growth variable
$g\equiv\delta/a$. Accuracies better than $0.05\%$ for $\Lambda$CDM
with $\Omega_{m0}\in[0.22,1]$ were reported, and $1\%$ when lowering
the limit to $\Omega_{m0}=0.01$. For $w\neq-1$, the paper reported
an accuracy within $0.4\%$ for $w=-0.5$ and for dynamical EoS $0.25\%$
when $w_{0}=-0.82$ and $w_{a}=0.58$. Here, we first will check how accurately
a constant $\gamma$ can reproduce $f$. In the next section, we will propose and
analyze a new fitting function for $\gamma$.

\begin{figure}[h]
\centering{}
\includegraphics[scale=0.6]{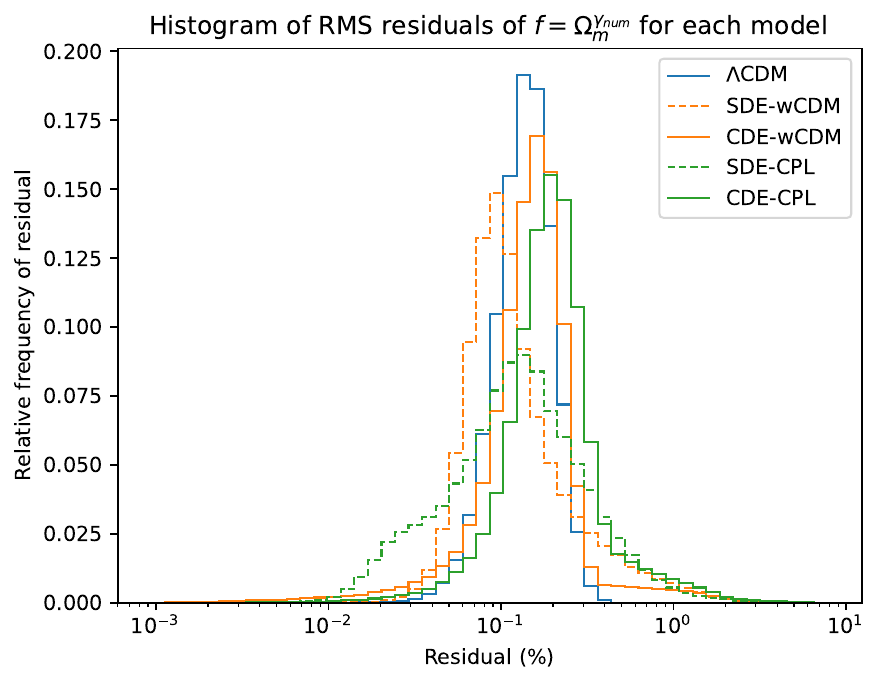}
\caption{\label{fig:hist-resid}
Histogram of RMS percent residuals for the constant growth index fit, with
respect to the numerical solution of $f$ for each model, normalized by relative
frequency.}
\end{figure}

In this work, we report the distribution of percent residuals given
by the Root Mean Square (RMS) of $1-\Omega_{m}^{\gamma_{{\rm
nun}}}\left(z\right)/f_{{\rm nun}}\left(z\right)$,
where $f_{{\rm nun}}$ is the numerical solution and $\gamma_{{\rm nun}}$
the corresponding fitted $\gamma$ value. We compute the RMS percent residuals
along $10$ evenly spaced values of $0\le z\le2$, and show its distribution in
Fig. \ref{fig:hist-resid} for all sets of parameters obtained in the MCMC analysis presented in Sect. \ref{sec:Results}. As can be seen, a very small
fraction of models have residuals greater than $0.5\%$. The worst
case occurs for CDE-CPL, reaching up to $4\%$, but still for a small
fraction of the realizations. In general, all models present a concentration
of residuals close to $0.15\%$. We also have checked that the largest
residuals are mainly associated with low $\Omega_{m0}$ values. This
demonstrates that, considering the CPL EoS, the constant $\gamma$
parametrization for the growth rate is very reliable, even in the
case of CDE. However, for more complex EoS parametrizations and for
realizations including a non-negligible DE density at high-$z$, the
constant-$\gamma$ parametrization might not be adequate \citep{Batista2014a}.

\section{\label{sec:Results} General distribution of $\gamma$}
\begin{figure}
\centering{}
\includegraphics[scale=0.6]{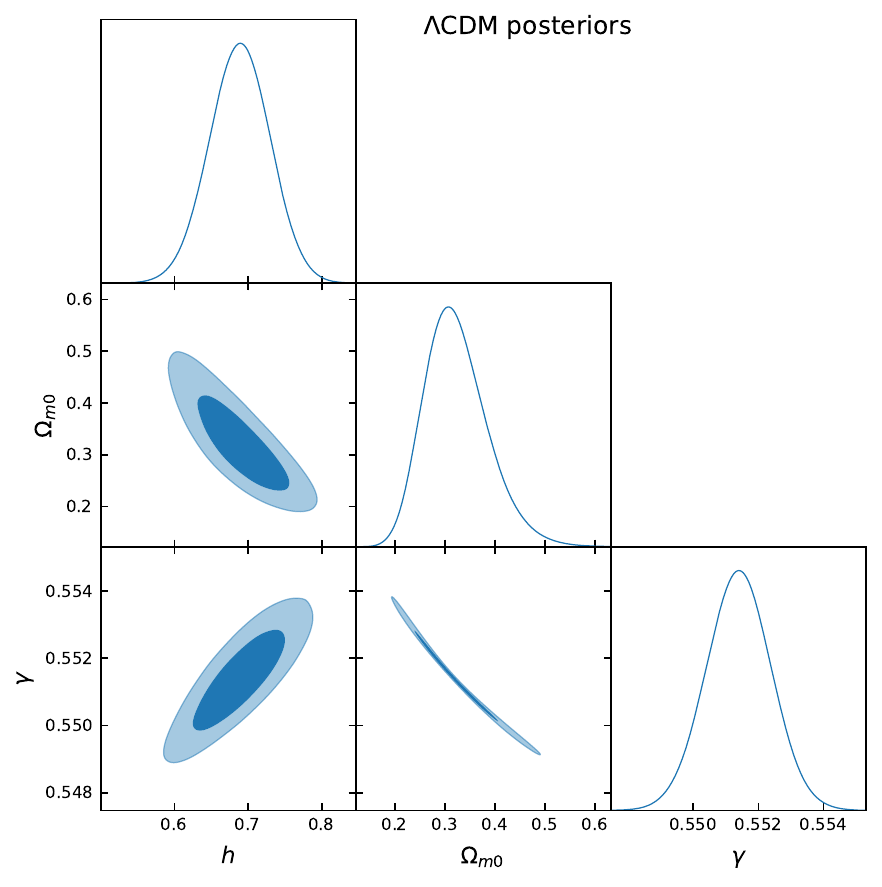}
\caption{\label{fig:trig-LCDM}
Marginalized posterior distributions for the $\Lambda$CDM background parameters, subjected to CC data and corresponding $\gamma$ values given by integration of Eq. (\ref{eq:delta-smooth-DE}).}
\end{figure}

\begin{figure*}
\centering{}
\includegraphics[scale=0.52]{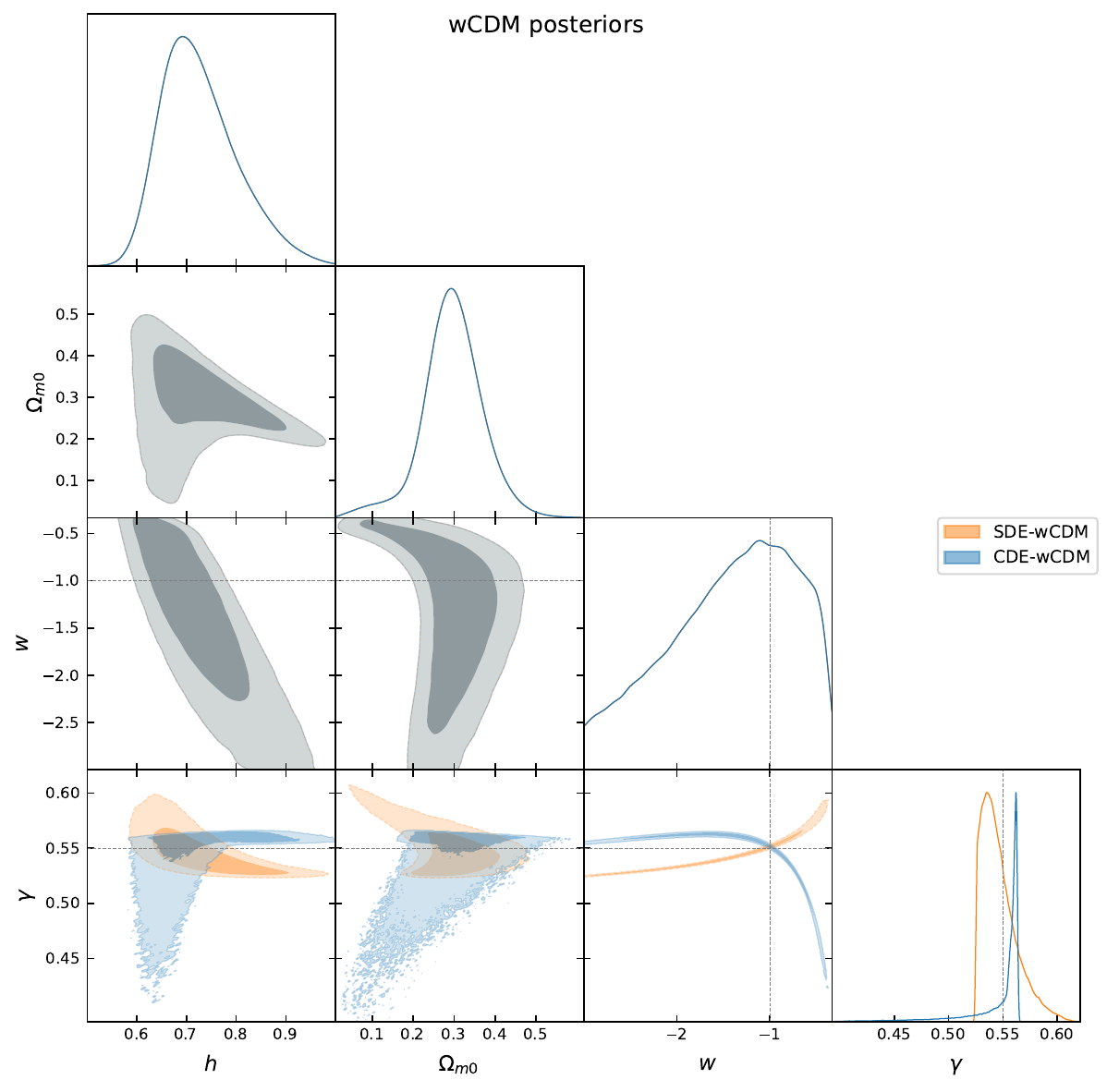}
\caption{\label{fig:trig-wCDM} Marginalized
posterior distributions for the $w$CDM models, with a shared background and
distinct Smooth and Clustering Dark Energy growth index distributions. The
dashed lines indicate the values of the $\Lambda$CDM parameters.}
\end{figure*}

\begin{figure*}
\centering{}
\includegraphics[scale=0.52]{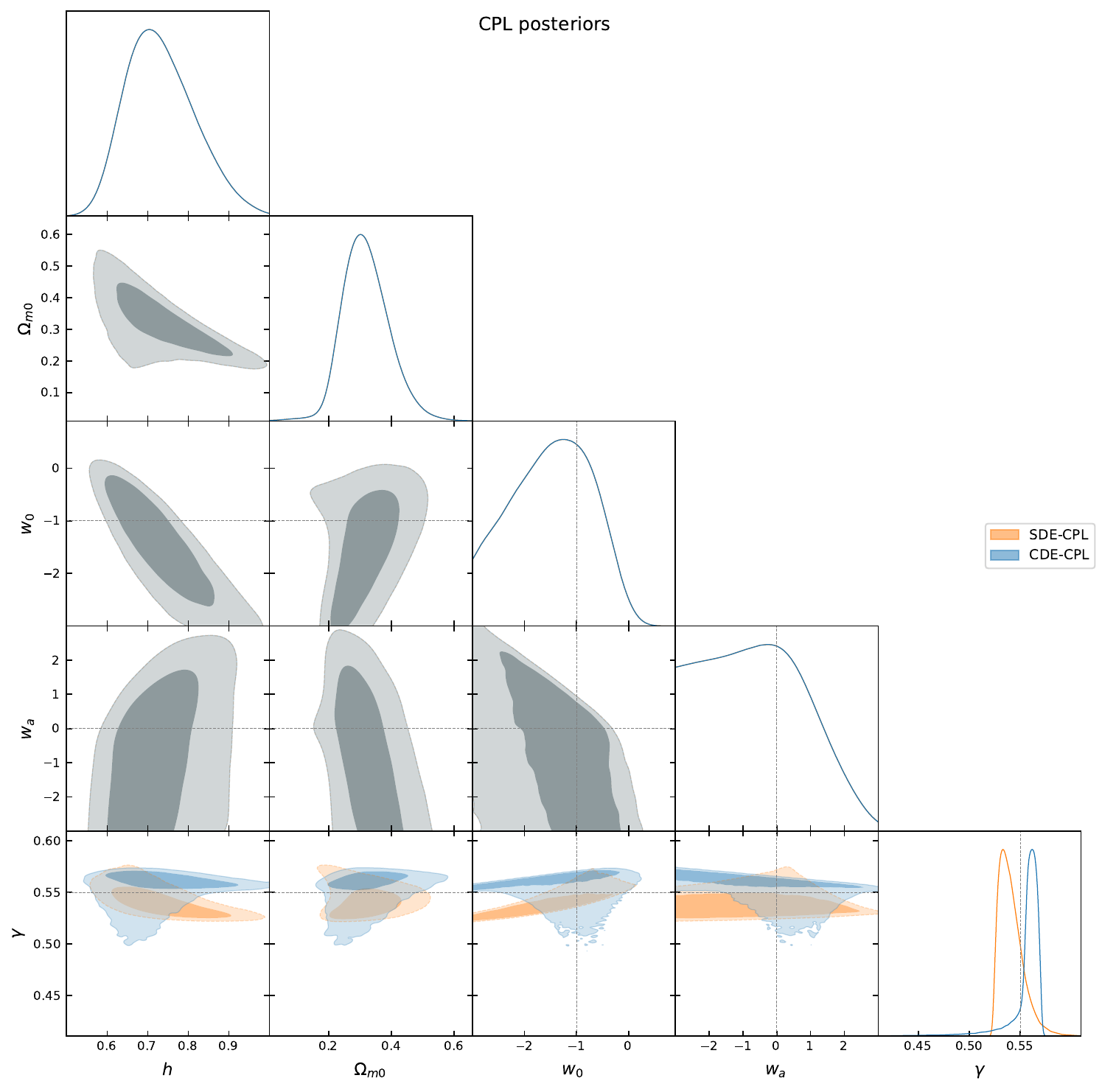}
\caption{\label{fig:trig-CPL}
Marginalized posterior distributions for the
CPL models, with a shared background and distinct Smooth and Clustering Dark
Energy growth index distributions. The dashed lines indicate the values of the
$\Lambda$CDM parameters.}
\end{figure*}

The parameter posteriors for $\Lambda$CDM, $w$CDM and CPL models are presented
in figures \ref{fig:trig-LCDM}, \ref{fig:trig-wCDM} and \ref{fig:trig-CPL},
respectively. Regarding the background parameters, we can see in Fig.
\ref{fig:trig-LCDM} that CC data can constrain the two free parameters,
$\Omega_{m0}$ and $h$, in the $\Lambda$CDM model. For $w$CDM this is not case,
mainly because $w$ is being limited by the prior assumed, but note that the
posterior also shows some preference for $w\simeq-1$.
Fig. \ref{fig:trig-CPL} also shows that CPL parameters, $w_{0}$
and $w_{a}$, can not be constrained by CC data. The posteriors of
$w_{0}$ and $w_{a}$ are dominated by the flat priors shown in Tab.
\ref{tab:Priors} together with the condition
$\Omega_{de}\left(z=1000\right)<0.01$. The mean values for the marginalized 1D
distributions are given in Tab. \ref{tab:posterior-fits}. The number of samples
in our chains is $N\sim1000\tau$, where $\tau$ is the largest auto-correlation
time of the parameters. This is well beyond the suggested $N>50\tau$
described in \citep{ForemanMackey2012} as convergence criteria.

We recall that the main purpose of this analysis is to provide a wide,
but still meaningful, distribution of background parameters that can
be used to compute the linear growth of matter perturbations and $\gamma$.
As can be seen in figures \ref{fig:trig-LCDM}, \ref{fig:trig-wCDM} and \ref{fig:trig-CPL}, despite this large variation of the EoS parameters,
$\gamma$ can be determined with much smaller variation. 

The $\Lambda$CDM posteriors shown in Fig. \ref{fig:trig-LCDM} display
a great determination of the growth index at
$\gamma_{\Lambda}=0.5514\pm0.0010$, agreeing with the previous results from
Refs. \citep{Wang:1998gt,Linder:2005in}. However, using data from Cosmological
Microwave Background, galaxy surveys and Baryon Acoustic Oscillation (BAO)
data, Ref. \citep{Nguyen:2023fip} found a value of
$\gamma _m=0.633_{-0.024}^{+0.025}$, excluding the $\Lambda$CDM value within
$3.7\sigma$, also showing that this value effectively solves the $S_{8}$
tension. Assuming normally distributed probabilities for the
measured value and the $\Lambda$CDM one, with mean and standard deviation given
in Tab. \ref{tab:posterior-fits}, a simple quantification of the tension between these values is given by
\begin{equation}
\# \sigma =\frac{|\gamma _m - \gamma_{\Lambda}|}{\sqrt{\sigma
_{m} ^2 +
\sigma _{\Lambda}^2}}=3.26\,
\label{eq:tension}
\end{equation}
where we used $\sigma _{m}=0.025$. If we assume the usual value
$\gamma_{\Lambda} = 0.55$ and neglect its uncertainty, we get $3.32\sigma$.

Now let us check whether more general SDE models and their CDE counterparts can
provide high $\gamma$ values. Although the analysis in Ref. \citep{Nguyen:2023fip} assumes the $\Lambda$CDM background to produce the $\gamma$ constraints, assessing the possible values of $\gamma$ in more general background
and perturbative models will help us to identify scenarios in which
$\gamma$ can be increased and how likely this can happen. In the next section, we will address the statistical compatibility between theoretical values $\gamma$ and its estimated value and uncertainties in models beyond $\Lambda$CDM.

We first analyze the correlations between the $\gamma$ and the
EoS parameters and how $\gamma$ changes with respect to the $\Lambda$CDM
value. The clearest case is for $w$CDM model, Fig. \ref{fig:trig-wCDM},
but the following analysis also holds for CPL model. For $w>-1$ (non-phantom
EoS) we have lower $\Omega_{m}\left(z\right)$ in the past with respect
to the $w=-1$ case. In the SDE scenario, this behavior causes a suppression
of growth, giving a higher $\gamma$. In the case of CDE, this correlation
is inverted because $w>-1$ induces positive $\delta_{de}$, which
will act as an extra source of the gravitational potential, enhancing
the growth and lowering $\gamma$. For $w<-1$ (phantom EoS) we have
higher $\Omega_{m}\left(z\right)$ in the past with respect to $w=-1$.
In SDE case, this induces a lower $\gamma$. Again, CDE inverts this
correlation because now negative $\delta_{de}$ is induced, reducing
the source of the gravitational potential, consequently increasing
$\gamma$. These correlations also hold for CPL, but since many realizations
shown in Fig. \ref{fig:trig-CPL} include transitions from phantom
to non-phantom EoS and vice-versa, they are not very clearly visualized.
However, in terms of $w_{1}$, these correlations become more evident
in the upper panels of Fig. \ref{fig:w-gamma}. Therefore, we can summarize the impact of DE model in the $\gamma$ , with respect to the $\Lambda$CDM value,
as follows:
\begin{itemize}
\item SDE, non-phantom EoS: $\gamma>0.55$
\item SDE, phantom EoS: $\gamma<0.55$
\item CDE, non-phantom EoS: $\gamma<0.55$
\item CDE, phantom EoS: $\gamma>0.55$
\end{itemize}

\begin{figure}
\begin{centering}
\includegraphics[scale=0.4]{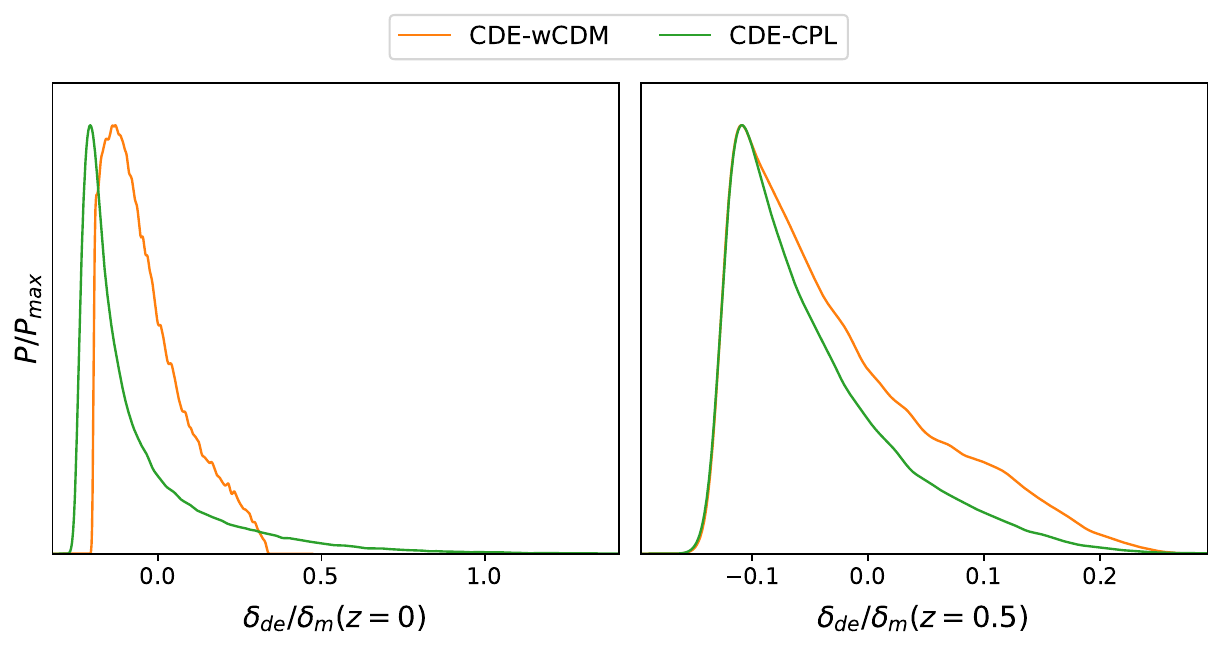}
\par\end{centering}
\centering{}\caption{\label{fig:delta-ratios}Distribution of perturbation ratios
for $w$CDM and CPL models at $z=0$ (left panel) and $z=0.5$ (right panel).}
\end{figure}

It is also interesting to check the frequency of positive or negative
occurrences of $\delta_{de}$ for the CDE scenario. In Fig.
\ref{fig:delta-ratios},
we present the distribution of $\delta_{de}/\delta_{m}$ at $z=0$
and $z=0.5$. We can see a small preference for negative $\delta_{de}$,
which is associated with the allowed values of the EoS parameters
as follows. For non-phantom EoS, $\Omega_{de}\left(z\right)$ can
be large at intermediate and high-$z$, a situation that is disfavored
by data, e.g., \citep{Gomez-Valent:2021cbe}. In our analysis, this
fact is mainly implemented with the prior $\Omega_{de}\left(z=1000\right)<0.01$.
On the other hand, for phantom EoS, $\Omega_{de}\left(z\right)$ is
very small at intermediate and high-$z$ and can still induce
an adequate accelerated expansion at low-$z$. Therefore, the allowed
parameter space for the EoS parameters has a larger fraction of phantom
realizations. As a consequence, given the correlations between $w$
and $\delta_{de}$ explained earlier, we see some preference for $\delta_{de}<0$.

\begin{figure}
\centering{}
\includegraphics[scale=0.58]{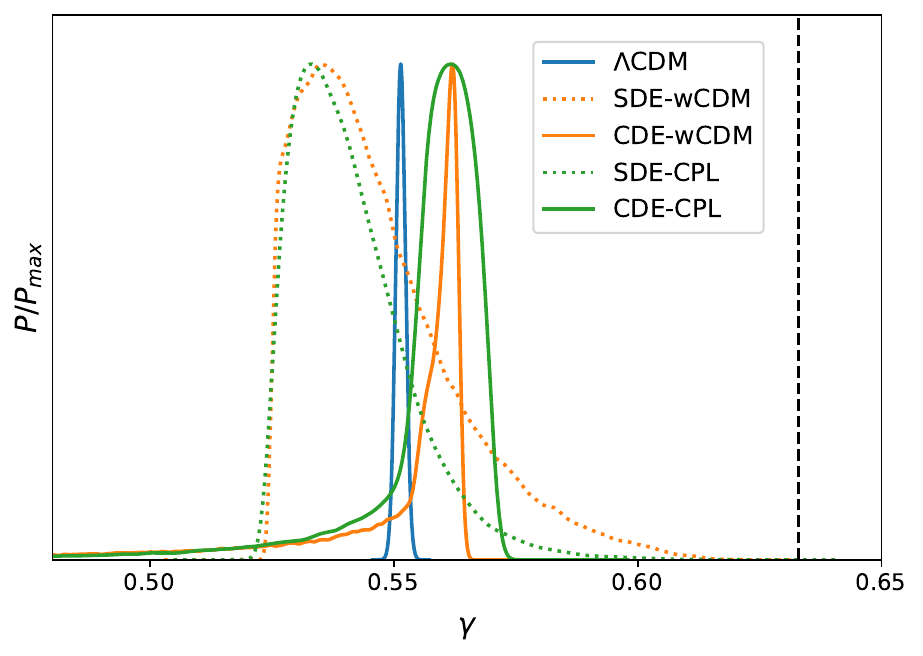}
\caption{\label{fig:dist-gamma}
Distribution for $\gamma$ in different models.
The vertical dashed line marks the measured value $\gamma=0.633$ for the $\Lambda$CDM background.}
\end{figure}

In Fig. \ref{fig:dist-gamma} we show a direct comparison between the $\gamma$
distributions obtained. As can be seen, the SDE models are similar, regardless
of the EoS parametrization. The same happens for the CDE models. Given the
larger fraction of phantom realizations, SDE models have a slight preference for
$\gamma<0.55$, whereas CDE prefers $\gamma>0.55$. The $\gamma$ distributions for
non-$\Lambda$ models have greater variance and asymmetry. However, none of these
models provide a significant fraction of realizations with $\gamma>0.6$.

Note that, in Fig. \ref{fig:dist-gamma}, the $\gamma$ distribution is heavy tailed for CDE models, then the reported mean value can lie outside
the $1\sigma$ interval reported in Tab. \ref{tab:posterior-fits}. We verified that the best fit model parameters lie within $1\sigma$ of the 1D marginalized distributions.

\begin{table}[h]
\caption{\label{tab:posterior-fits}Marginalized 1D constraints and $68\%$
C.L. intervals for the parameters of each cosmology subjected to CC data.}

\centering{}%
\begin{tabular}{cccc}
Parameter & $\Lambda$CDM & $w$CDM & CPL\tabularnewline
\hline 
\hline 
$h$ & $0.689\pm0.041$ & $0.727_{-0.092}^{+0.054}$ &
$\ensuremath{0.733_{-0.10}^{+0.069}}$\tabularnewline
$\Omega_{m0}$ & $0.324_{-0.072}^{+0.049}$ & $0.296\pm0.076$ &
$0.319_{-0.082}^{+0.060}$\tabularnewline
\hline 
$w$ or $w_{0}$ & - - - & $-1.42_{-0.37}^{+0.97}$ &
$-1.47_{-0.72}^{+0.91}$\tabularnewline
$w_{a}$ & - - - & - - - & $-0.6_{-2.0}^{+1.1}$\tabularnewline
\hline 
$\gamma$ (SDE) & $0.5514\pm0.0010$ & $0.5465_{-0.021}^{+0.0054}$ &
$0.5412_{-0.015}^{+0.0056}$\tabularnewline
$\gamma$ (CDE) & - - - & $0.545_{+0.0061}^{+0.019}$ &
$0.556_{-0.00039}^{+0.013}$\tabularnewline
\hline 
\end{tabular}
\end{table}

We could expect that phantom CDE models are able to produce high $\gamma$
values because, as explained earlier, the associated negative $\delta_{de}$
operates in this direction. However, given that the gravitational
potential depends on $\Omega_{de}\delta_{de}$ and, in the phantom
case, $\Omega_{de}$ is small at intermediate and high-$z$, the actual
impact of these models on $\gamma$ is very limited. Consequently,
although some preference for $\gamma>0.55$ can be seen in Fig.
\ref{fig:dist-gamma},
the highest possible values virtually never reach $\gamma\gtrsim0.58$.

\subsection*{New fits for $\gamma(w)$}

With the posterior distributions obtained, we were able find a more
accurate and general fitting function for $\gamma$, given by
\begin{equation}
\gamma(w_{1})=aw_{1}+be^{cw_{1}}+d,\label{eq:fitting-func}
\end{equation}
The coefficients for each model are listed in Tab. \ref{tab:fits-coefficients}.
The tested range of $w_{1}$ is $[-3,-1/3]$. The main difference
between SDE and CDE models is the sign of the coefficient in the exponential
term.

\begin{figure*}[t]
\begin{centering}
\includegraphics[scale=0.56]{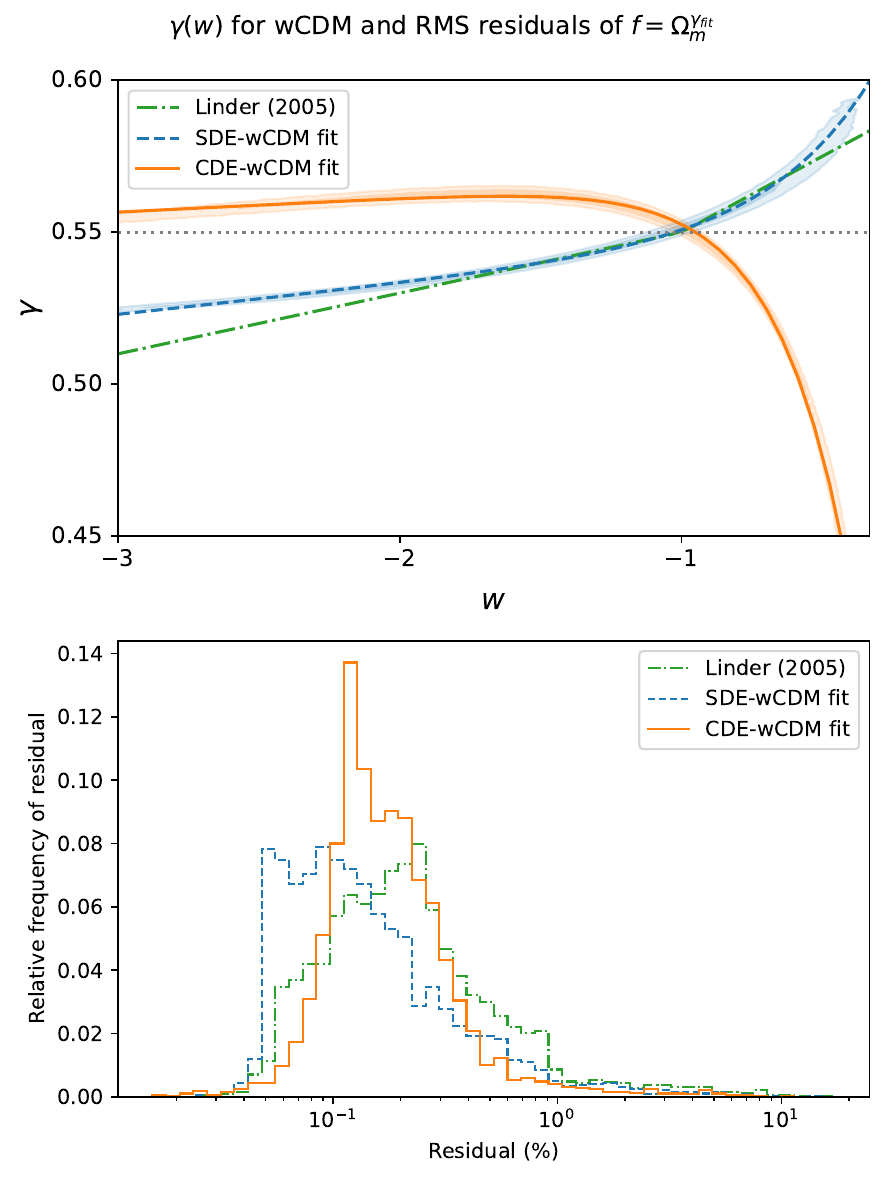}
\includegraphics[scale=0.56]{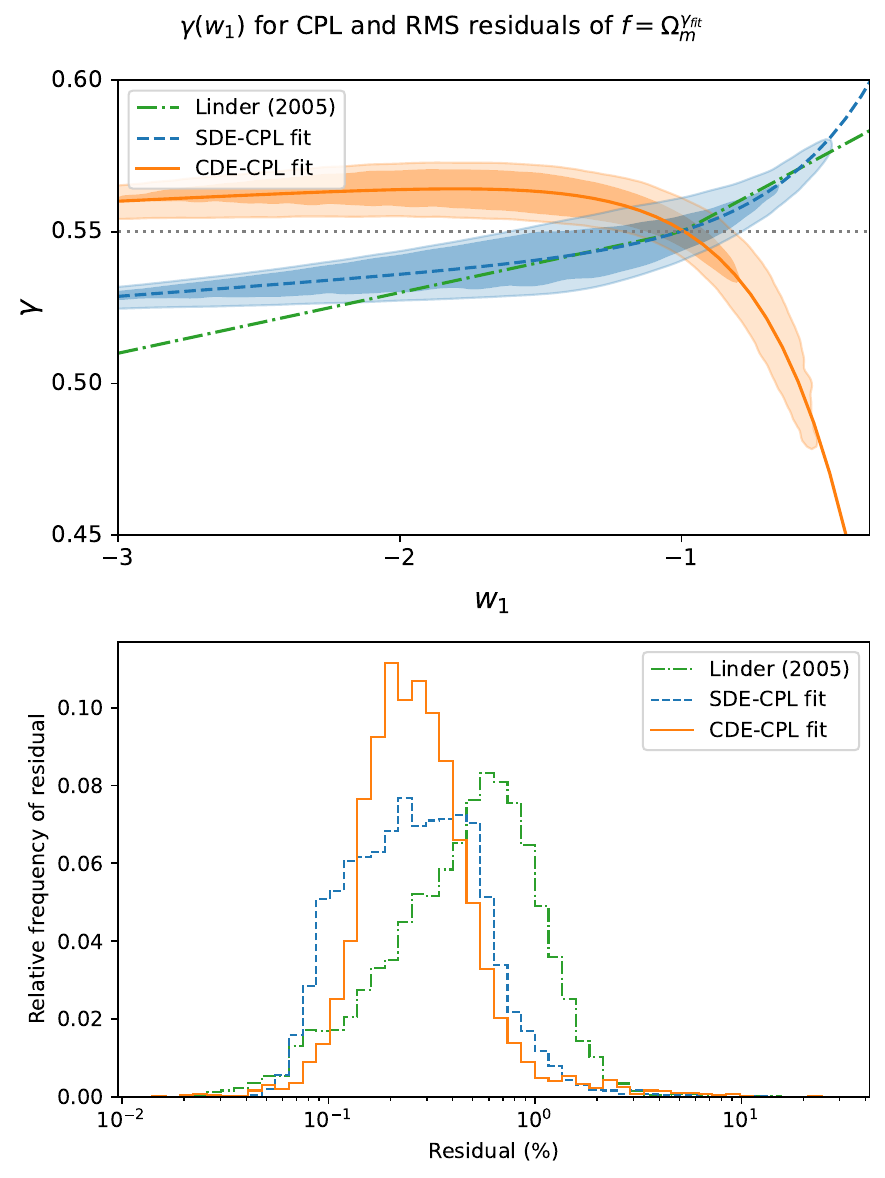}
\par\end{centering}
\caption{\label{fig:w-gamma}Different fits of $\gamma(w_{1})$ for $w$CDM
models (left panels) and for CPL models (right panels) with contours
for $\gamma(w_{1})$ for SDE (blue) and CDE (orange) in the the upper
panels. The lower panels show histograms, normalized by relative frequency,
of the RMS percent residual between the $\gamma$ given by fitting
functions and the corresponding numerical value, obtained from the
parametrization $f=\Omega_{m}^{\gamma}$. }
\end{figure*}

\begin{table}[h]
\begin{centering}
\caption{\label{tab:fits-coefficients}Coefficients for the fits of
$\gamma(w_{1})=aw_{1}+be^{cw_{1}}+d$
for each model.}
\begin{tabular}{cccccc}
Model & Cosmology & $a$ & $b$ & $c$ & $d$\tabularnewline
\hline 
\hline 
SDE & $w$CDM & $0.0101$ & $0.1266$ & $2.8013$ & $0.5532$\tabularnewline
SDE & CPL & $0.0068$ & $0.1346$ & $2.8453$ & $0.5493$\tabularnewline
CDE & $w$CDM & $0.0047$ & $-0.6320$ & $3.8570$ & $0.5706$\tabularnewline
CDE & CPL & $0.0045$ & $-0.4673$ & $3.2250$ & $0.5738$\tabularnewline
\hline 
\end{tabular}
\par\end{centering}

\end{table}

The quality of the fits are demonstrated in the upper panels of Fig.
\ref{fig:w-gamma}, where our fits are well contained within the posteriors
contours. The lower panels show the residuals of the parametrization
of Eq. (\ref{eq:gamma-linder}) and ours. We computed these residuals with the
same RMS definition used for Fig. \ref{fig:hist-resid}. 

Let us first analyze the results for SDE in Fig. \ref{fig:w-gamma}.
Our fit is most accurate for the $w$CDM model (left panels), presenting
a mode of residuals at $0.1\%$, while the linear fit of Eq.
(\ref{eq:gamma-linder})
presents a mode around $0.25\%$. In any case, very few realizations
have residuals $>1\%$. For the CPL models (right panels), our
fit is noticeably better than the one of Eq. (\ref{eq:gamma-linder}).
For instance, the linear fit residuals, Eq. (\ref{eq:gamma-linder}),
peaks around $1\%$ while the exponential fit, Eq. \eqref{eq:fitting-func},
peaks at $0.2\%$. The main reason for this improvement is that the linear
fit does not perform well for $w_{1}<-2$, as can be seen in the top
panels of Fig. \ref{fig:w-gamma}. We have tested whether there is some
improvement in the fits quality when changing the pivot $w(z=1)$ to other values. For 
$w(z=1.2)$ and $w(z=0.8)$, we noticed no relevant modifications in the histograms
of residuals.

Our fitting function for $\gamma$ can also be used to describe the
values for the CDE case, which are shown with orange lines and contours
in Fig. \ref{fig:hist-resid}. As can be seen, it also performs
very well in models with clustering DE. The modes of the RMS residuals
is around $0.1\%$ for both $w=\text{const.}$ and CPL parametrizations.

\section{\label{estimation} Estimating the $\gamma$ tension beyond $\Lambda$CDM}
So far, we have analyzed the distribution of $\gamma$ based on the very loose constraints on background evolution given by the CC data. This study was helpful to analyse the accuracy of the constant $\gamma$ parametrization, the possible $\gamma$ values and determine more general and accurate fitting functions. Now we turn our attention to whether DE models have any significant tension with the high values of $\gamma$, as suggested by measurement for $\Lambda$CDM. As shown in the previous section, both SDE and CDE can not provide a significant fraction realizations with $\gamma\simeq 0.63$. However, it is important to note this $\gamma$ value was determined assuming the $\Lambda$CDM background, and a direct comparison between the measurement and possible values in other models might be inconsistent. 

\begin{figure*}[t]
\begin{centering}
\includegraphics[scale=0.5]{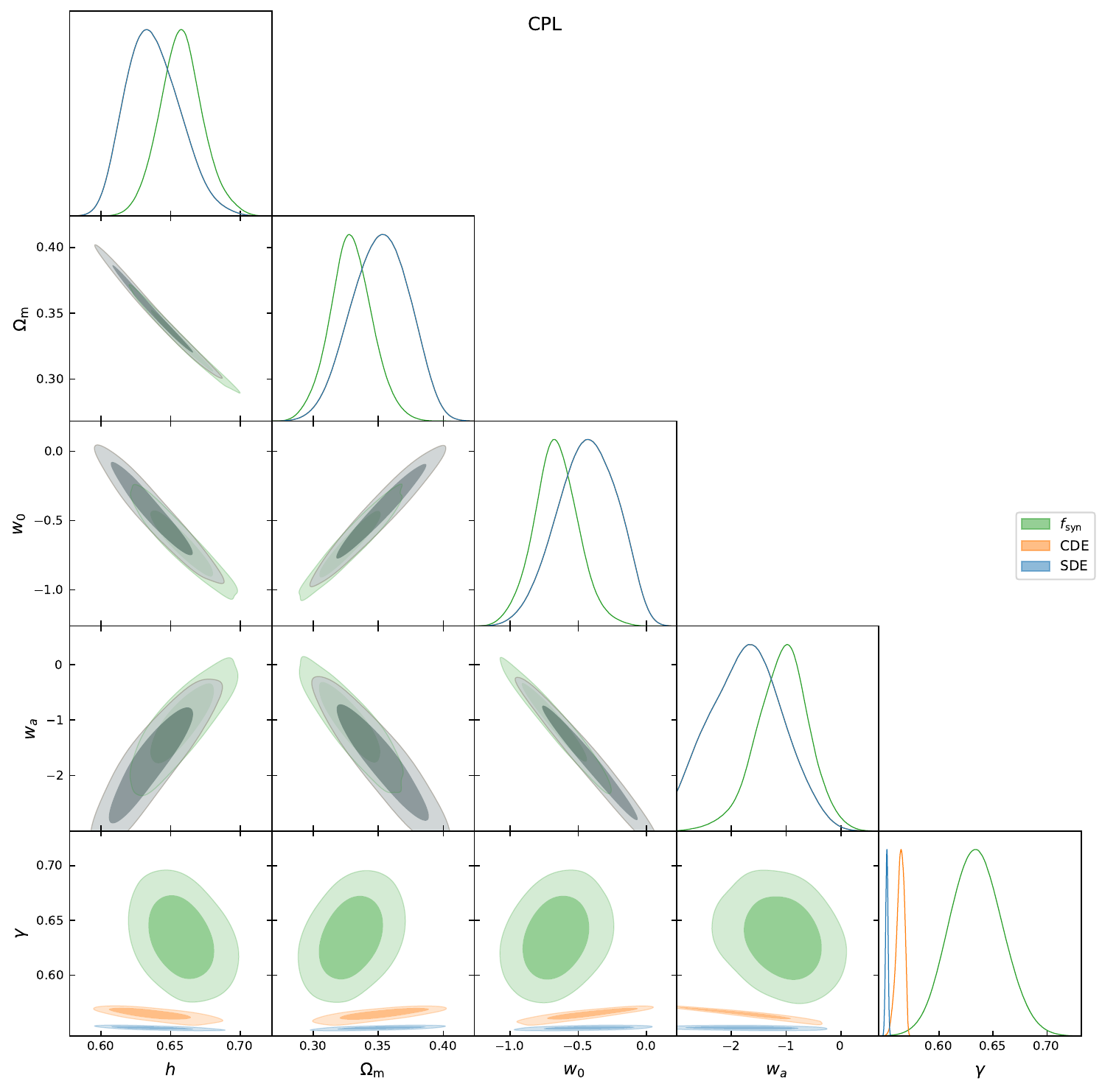}
\par\end{centering}
\caption{Marginalized posterior distributions for the CPL background: constrained by DESI BAO DR2 + $(\theta_*, \omega_b, \omega_{bc})_{\rm CMB}$ and corresponding theoretical $\gamma$ values for SDE and CDE models and the estimated constraints on $\gamma$ ($f_{\rm syn}$) using DESI BAO DR2 + $(\theta_*, \omega_b, \omega_{bc})_{\rm CMB}$ + $f_{\rm syn}$ data. Note that SDE and CDE models share the same distributions for background parameters.
\label{fig:trig_CPL-DESI}}
\end{figure*}

In order to make more precise estimates, in this section, we constrain the background parameters using DESI DR2 BAO together with CMB correlated priors on $(\theta_*, \omega_b, \omega_{bc})_{\rm CMB}$, where $\omega_{b}=\Omega_{b0} h^2$ is the baryonic physical density parameter, $\omega_{bc}=(\Omega_{b0}+ \Omega_{c0}) h^2$ is the sum of the previous quantity and the physical cold dark matter density parameter and $\theta_*$ is the angular scale of the CMB acoustic peaks, as discussed in Ref. \cite{DESI:2025zgx}. We use the same set of priors indicated in \cite{DESI:2025zgx}. For this analysis and following ones, we make use of \texttt{CAMB} \cite{Lewis2000, Howlett:2012mh} to compute the BAO observables and the \texttt{Cobaya} MCMC sampler \cite{Torrado2020, Lewis:2002ah}, which has the likelihood for DESI DR2 BAO implemented. As a convergence criteria for the chains, we assume the Gelman-Rubin statistic satisfying $R-1<0.01$. 

In Tab. \ref{tab:DESI-posteriors-all}, we show the marginalized (68\% C.L.) constraints on the background parameters and the resulting theoretical values for $\gamma$. As can be seen, all parameters now have much smaller uncertainties with respect to the analysis using CC data, shown in the previous section, Tab. \ref{tab:posterior-fits}.


To make a fast estimation of the $\gamma$ values and its uncertainty in $w$CDM and CPL backgrounds, we proceed as follows. Assuming fiducial parameters $\Omega_{m0}=0.3$, which is the visualized central value of the most precise measurement in Ref. \citep{Nguyen:2023fip} for $\Lambda$CDM, we generate a synthetic growth rate function based on the measured $\gamma$, 
\begin{equation}
f_{\rm syn}(z)=\Omega_{m\Lambda}^{0.633}(z)\,.
\label{f-syn}
\end{equation}
We use this function to generate $18$ mock data points of $f_{\rm syn}(z)$, evenly spaced in the range $0\le z \le 0.9$ with normally distributed $7\%$ scatter around the fiducial values. Then we use the mock data points in an uncorrelated Gaussian likelihood function for $f_{\rm syn}(z)$. 

With this setup, realizations of $f_{\rm syn}(z)$ with $7\%$ uncertainty converge to reproduce the measured constraints on $\gamma$ obtained in Ref. \cite{Nguyen:2023fip}. We choose a particular realization that yields constraints shown in the middle section of Tab. \ref{tab:DESI-posteriors-all}, in particular, $\gamma=0.625\pm0.023$ for $\Lambda$CDM. Note that, as stated in the DESI BAO DR2 analysis, Ref. \cite{DESI:2025zgx}, these measurements are based on a dataset that is essentially background-dependent and model-independent. 

\begin{figure}[]
\centering{}
\includegraphics[scale=0.58]{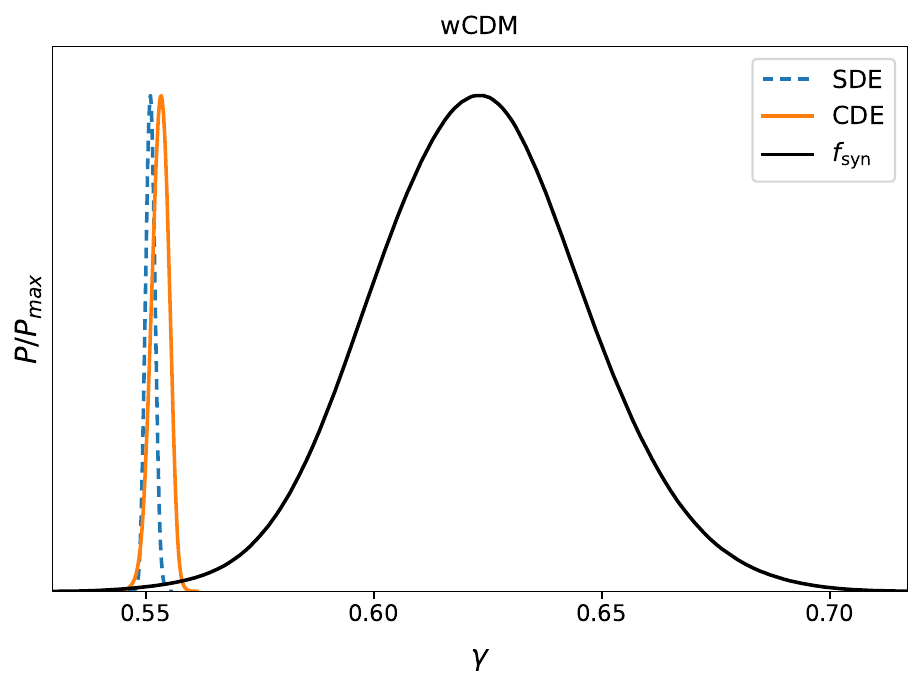}
\caption{\label{fig:f_synth_wCDM}
Distribution for $\gamma$ in $w$CDM model for SDE and CDE and the estimated constraints on $\gamma$ based on $f_{\rm syn}$ data with $7\%$ uncertainty.}
\end{figure}

\begin{figure}[]
\centering{}
\includegraphics[scale=0.58]{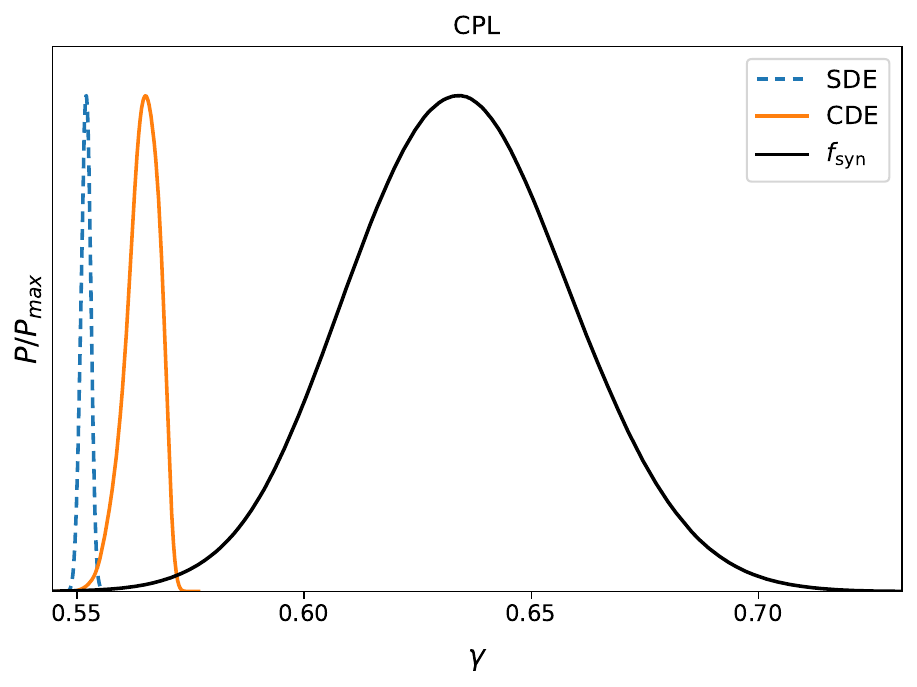}
\caption{\label{fig:f_synth_CPL}
Distribution for $\gamma$ in CPL model for theoretical SDE and CDE models and estimated constraints of $\gamma$ based on $f_{\rm syn}$ data with $7\%$ uncertainty.}
\end{figure}

For $w$CDM, as can be seen in Tab. \ref{tab:DESI-posteriors-all} and in Fig. \ref{fig:f_synth_wCDM}, the central value of the estimated $\gamma$ is very close to the measured one for $\Lambda$CDM. For CPL, there is a small increase in the estimated $\gamma$, shown in Tab. \ref{tab:DESI-posteriors-all} and in Fig. \ref{fig:f_synth_CPL}. In both cases, the uncertainty in $\gamma$ is close to the one found for $\Lambda$CDM. This finding is somewhat unexpected because usually measurements for models with larger parameter space are more inaccurate. We checked that this actually occurs when using Gaussian uncorrelated priors based on the posteriors obtained for the background parameters. Moreover, we also tested the estimation of $\gamma$ constraining the background with CC data. In this case, given the much larger allowed parameter space volume, we also see a relevant increase in the $\gamma$ uncertainty for $w$CDM and CPL models. Finally, we checked that less precise mock data, yielding $0.028$ uncertainty for $\gamma$ in $\Lambda$CDM model, support the same conclusions of nearly unchanged  $\gamma$ central values and uncertainties for the other backgrounds when using DESI BAO DR2 + $(\theta_*, \omega_b, \omega_{bc})_{\rm CMB}$. Therefore, we understand that the same level of uncertainty for $\gamma$ for all backgrounds is due to the high constraining power and strong correlations in the background parameters determined by DESI BAO DR + $(\theta_*, \omega_b, \omega_{bc})_{\rm CMB}$. 

As a visual example of these correlations, some of which were already shown in Ref. \cite{DESI:2025zgx}, in Fig. \ref{fig:trig_CPL-DESI}, we show the posteriors for the CPL model parameters. We can also see that the inclusion of $f_{\rm syn}$ data has some impact on the central value of background parameters and decreases their uncertainties. In particular, the determination of $w_a$ has a relevant improvement in its determination, excluding the lowest values permitted by the background analysis. In this figure, note that SDE and CDE have the same background posteriors.       

In Figs. \ref{fig:f_synth_wCDM} and \ref{fig:f_synth_CPL}, we show the theoretical predictions for $\gamma$ in $w$CDM and CPL models and $\gamma$ estimated measurements. As can be seen, for $w$CDM, the central value is basically the same for SDE and CDE. This is expected because the posterior of $w$ gives $w\simeq-1$ with small uncertainties, making CDE effects much smaller than those seen in Fig. \ref{fig:dist-gamma}. In the CPL case, we see a more significant difference between the central $\gamma$ values for SDE and CDE, with the latter giving higher values. As explained earlier, this is a manifestation of the preference for phantom EoS (at least for some period of the cosmic expansion), which is favored by DESI data, mainly due to preference for negative $w_a$.  

In the present analysis, the theoretical $\gamma$ distributions are much less asymmetric than the more general distributions shown in Sect. \ref{sec:Results}. Therefore, we can more safely estimate the level of tension, as defined in Eq. (\ref{eq:tension}) for Gaussian distributions. These results are shown in the bottom section of Tab. \ref{tab:DESI-posteriors-all}. For the CDE cases, which still have slightly asymmetric $1\sigma$ limits, we use the larger value. As can be seen, the tension is similar in all models, except for smooth CPL, which has a slightly increased tension with the estimated $\gamma$. This occurs because the central value of $\Omega_{m0}$ for the CPL background has a significant increase with respect to the $\Lambda$CDM or $w$CDM values. As a consequence, in order to have a similar growth rate, $\gamma$ has to increase. Since the smooth CPL model has a very limited variation of $\gamma$, around the $\Lambda$CDM value, the tension increases. On the other hand, given the perturbative behavior already explained, the clustering CPL model has more freedom to raise the $\gamma$ value, resulting in a lower tension with respect to its smooth counterpart.

%

Therefore, also when constraining the background parameters with DESI+$(\theta_*, \omega_b, \omega_{bc})_{\rm CMB}$, neither SDE nor CDE with CPL EoS can raise the mode of the their $\gamma$ distributions to high values, $\gamma>0.6$. Moreover, our method to estimate $\gamma$ for $w$CDM and CPL backgrounds does not indicate any significant variation of it with respect to the measured value for the $\Lambda$CDM model and quite similar uncertainties, which is likely caused by the highly constrained background parameters and their strong correlations. 

If future analysis confirm this high value for $w$CDM and CPL backgrounds, DE models will be challenged. Furthermore, it is interesting to highlight that, given that CPL background subjected to DESI+$(\theta_*, \omega_b, \omega_{bc})_{\rm CMB}$ data has higher $\Omega_{m0}$, which enhances the growth rate, there is a natural tendency for higher $\gamma$, which counteracts this enhancement. Thus, it is quite likely that $\gamma$ tension can not be alleviated by smooth CPL models. However, the concrete tension level crucially depends on $\gamma$ measurements with real data.

\begin{table*}[t]
\caption{\label{tab:DESI-posteriors-all} Upper section: marginalized (68\% C.L.) constraints on the background parameters using DESI + $(\theta_*, \omega_b, \omega_{bc})_{\rm{CMB}}$ data and the resulting theoretical $\gamma$ constraints.  Middle section: marginalized (68\% C.L.) constraints on parameters using DESI+$(\theta_*, \omega_b, \omega_{bc})_{\rm{CMB}}$+$f_{\rm syn}$ data and the corresponding estimated $\gamma$ constraints for each background. Bottom section: tension between the theoretical and estimated $\gamma$ values.}

\centering{}
\begin{tabular}{cccc}
Dataset/Parameter & $\Lambda$CDM & $w$CDM & CPL\tabularnewline
\hline
\hline
\multicolumn{4}{l}{\bf DESI-DR2+$(\theta_*, \omega_b, \omega_{bc})_{\rm CMB}$}\tabularnewline
$h$            & $0.6836_{-0.0031}^{+0.0028}$ & $0.6904\pm0.0097$ & $0.638_{-0.022}^{+0.017}$\tabularnewline
$\Omega_{m0}$  & $0.3005\pm0.0038$ & $0.2957\pm0.0076$ & $0.351\pm0.022$\tabularnewline
$w$ or $w_{0}$ & - - - & $-1.029\pm0.039$ & $-0.44\pm0.22$\tabularnewline
$w_{a}$        & - - - & - - - & $-1.67\pm0.64$\tabularnewline
$\gamma$ (SDE) & $0.551678_{-0.000070}^{+0.000059}$ & $0.5509\pm0.0010$ & $0.5519\pm0.0011$ \tabularnewline
$\gamma$ (CDE) & - - - & $0.5530_{-0.0017}^{+0.0021}$ & $0.5644_{-0.0028}^{+0.0043}$ \tabularnewline
\hline
\multicolumn{4}{l}{\bf DESI-DR2+$(\theta_*, \omega_b, \omega_{bc})_{\rm CMB}$+$f_{\rm syn}$} \tabularnewline
$h$            & $0.6834\pm0.0029$ & $0.6853^{+0.0082}_{-0.0092}$ & $0.647\pm0.011$ \tabularnewline
$\Omega_{m0}$  & $0.3008\pm0.0038$ & $0.2994\pm0.0070$ & $0.340\pm0.012$ \tabularnewline
$w$ or $w_{0}$ & - - - & $-1.008^{+0.037}_{-0.033}$ & $-0.54\pm0.12$ \tabularnewline
$w_{a}$        & - - - & - - - & $-1.41\pm0.36$ \tabularnewline
$\gamma$ ($f_{\rm syn}$) & $0.625\pm0.023$ & $0.624\pm0.024$ & $0.639\pm0.024$ \tabularnewline
\hline
\multicolumn{4}{l}{\bf Tension}\tabularnewline
$\#\sigma$ ($f_{\rm syn}$ - SDE) & $3.06$ & $3.04$ & $3.63$ \tabularnewline
$\#\sigma$ ($f_{\rm syn}$ - CDE) & - - -  & $2.95$ & $3.06$ \tabularnewline
\hline
\end{tabular}
\end{table*}

\section{\label{sec:Conclusion} Conclusions}
In this paper, we have analyzed the possibilities of DE models described by the CPL parametrization of producing a high $\gamma$ values, in line with $\gamma = 0.633$ obtained in Ref. \cite{Nguyen:2023fip} for $\Lambda$CDM. As summarized in Fig. \ref{fig:dist-gamma}, both smooth DE and clustering DE have $\gamma$ distributions with central values below $0.56$. This result depends
only on the following assumptions: (1) $H\left(z\right)$ must be compatible with
CC data, (2) $\Omega_{de}\left(z=1000\right)<0.01$ and (3) the priors described
in Tab. \ref{tab:Priors}. The combination of these assumptions
imposes very loose constraints on $w_{0}w_{a}$ parameters, allowing for a vast,
but meaningful, exploration of the possible values of $\gamma$. 

We analyzed the correlations between $\gamma$, the EoS parameters and
the clustering properties of DE. In particular, we can expect that phantom CDE
models can raise the value of $\gamma$. However, since the impact of such models
depend on $\Omega_{de}\left(z\right)\delta_{de}(z)$, the actual change is very
limited because $\Omega_{de}\left(z\right)$ decays rapidly with $z$. Non-phantom
SDE can also give higher $\gamma$, but these realizations are correlated with
very low matter density parameter, $\Omega_{m0}\simeq 0.1$.

We also proposed a new fitting function for
$\gamma=\gamma(w_{1})$, with overall accuracies better than the linear fit of Eq. (\ref{eq:gamma-linder}) covering the interval of $-3\le w_{1}\le-1/3$. For the first time, we produced a $\gamma$ fitting function for CDE models, Eq. (\ref{eq:fitting-func}) with coefficients given in Tab. \ref{tab:fits-coefficients}. These fits can be useful for a fast determination of $\gamma$ for both SDE and CDE models described by CPL parametrization.

In a more specific analysis, we have computed the theoretical $\gamma$ distributions constraining the background parameters using DESI BAO DR2 + $(\theta_*, \omega_b, \omega_{bc})_{\rm CMB}$ data. Also in this case, the $\gamma$ values are not allowed to reach high values, obtaining central values with $\gamma\lesssim0.57$. In the same context for background constraints, we also estimated the $\gamma$ values for $w$CDM and CPL models based on mock data generated for the growth rate function. This analysis indicated no significant change in the estimated $\gamma$ distribution and its uncertainty with respect to the measured value for $\Lambda$CDM. Consequently, this preliminary study suggests that the DE models considered in this work can have roughly the same level of $\gamma$ tension as found for $\Lambda$CDM, with a small tension decrease for CDE models relative to the SDE ones.

If the high $\gamma$ value found in Ref. \citep{Nguyen:2023fip} is validated by other observations and analysis \footnote{A few days after the first version of this paper was posted on arxiv.org, Ref. \cite{Specogna:2024euz} claimed that the use of more recent likelihoods for Planck data provide $\gamma$ measurements compatible with $\Lambda$CDM.} and for backgrounds beyond $\Lambda$CDM, it poses a significant challenge to DE models based on minimally coupled scalar fields that can be described by the CPL EoS and a constant $c_s$ on small scales because these models are incapable to provide such high $\gamma$ values. Since we have considered the two limiting cases of smooth DE ($c_s=1$) and full clustering DE ($c_s=0$), it seems unlikely that intermediate or
time-varying $c_s$ values can produce a significantly higher $\gamma$. A
possible alternative is to consider more general EoS parametrizations. However,
as demonstrated in our study, large variations on the EoS parameters and
$\Omega_{m0}$ produce much narrower $\gamma$ distributions. Therefore, in
principle, even more general EoS parametrizations should have some difficulties
in explaining such high $\gamma$ values, with the disadvantage of introducing
more parameters. If this `$\gamma$ tension' persists, we might be seeing an
early evidence of modified gravity or non-standard dark matter.

\begin{acknowledgments}
We thank Valerio Marra for useful discussions. IBSC thanks Universidade Federal
do Rio Grande do Norte for the undergraduate research scholarship, N° 01/2023
(PIBIC-PROPESQ) project PIJ20915-2023.
\end{acknowledgments}

\bibliography{referencias.bib}

\end{document}